\documentclass[fleqn,twoside]{article}
\usepackage{espcrc2}

\usepackage{graphicx}
\usepackage[figuresright]{rotating}

\newcommand{\AmS}{{\protect\the\textfont2
  A\kern-.1667em\lower.5ex\hbox{M}\kern-.125emS}}

\def\Journal#1#2#3#4{{#1} {\bf #2} (#3) #4}

\def\NP{\em Nucl. Phys.}

\def\PL{\em Phys. Lett.}

\def\JP{\em J. Phys.}
\def\EPJ{\em Eur. Phys. J.}
\def\EPJD{\em Eur. Phys. J. direct}

\def\JHEP{\em JHEP}
\def\etal{{\it et al.}}

\def\be{\begin{equation}}
\def\ee{\end{equation}}
\def\bea{\begin{eqnarray}}
\def\eea{\end{eqnarray}}

\newcommand{\dd}       {\mathrm{d}}

\newcommand{\calP}       {\ensuremath{\mathcal{P}}}

\newcommand{\bbbar}     {\ensuremath{\mathrm{b\bar{b}}}}

\newcommand{\epem}              {\ensuremath{\mathrm{e^+e^-}}}

\newcommand{\CA}                {\ensuremath{C_A}}
\newcommand{\CF}                {\ensuremath{C_F}}

\newcommand{\nf}                {\ensuremath{n_f}}
\newcommand{\as}                {\ensuremath{\alpha_\mathrm{S}}}
\newcommand{\asrs}                {\ensuremath{\alpha_\mathrm{S}(\sqrt{s})}}

\newcommand{\asmu}              {\ensuremath{\alpha_\mathrm{S}(\mu)}}

\newcommand{\asmz}              {\ensuremath{\alpha_\mathrm{S}(M_{\mathrm{Z^0}})}}
\newcommand{\azmui}               {\ensuremath{\alpha_0(\mu_\mathrm{I})}}

\newcommand{\az}               {\ensuremath{\alpha_0}}
\newcommand{\mui}               {\ensuremath{\mu_\mathrm{I}}}

\newcommand{\aztwo}    {\ensuremath{\alpha_0(\mathrm{2\,GeV})}}

\newcommand{\oaa}               {\ensuremath{\mathcal{O}(\alpha_\mathrm{S}^2)}}

\newcommand{\znull}     {\ensuremath{\mathrm{Z^0}}}

\newcommand{\mz}                {\ensuremath{M_{\mathrm{Z^0}}}}

\newcommand{\var}       {\ensuremath{\mathcal{F}}}

\newcommand{\cp}                {\ensuremath{C}}
\newcommand{\bt}                {\ensuremath{B_\mathrm{T}}}
\newcommand{\bw}                {\ensuremath{B_\mathrm{W}}}
\newcommand{\mh}                {\ensuremath{M_\mathrm{H}}}

\newcommand{\thr}               {\ensuremath{1-T}}

\newcommand{\chisq}     {\ensuremath{\chi^2}}
\newcommand{\chisqd}    {\ensuremath{\chi^2/\mathrm{d.o.f.}}}
\newcommand{\xmu}               {\ensuremath{x_{\mu}}}

\newcommand{\lnr}               {\ensuremath{\ln(R)}}

\newcommand{\yy}                {\ensuremath{y_{23}}}

\newcommand{\rs}                {\ensuremath{\sqrt{s}}}

\newcommand{\gev}               {\ensuremath{\mathrm{GeV}}}

\newcommand{\py}                {{\sc Pythia}}
\newcommand{\hw}                {{\sc Herwig}}
\newcommand{\ar}                {{\sc Ariadne}}
\newcommand{\jt}                {{\sc Jetset}}
\newcommand{\cj}                {{\sc Cojets}}

\title{\as\ and Power Corrections from JADE
\vspace*{-35mm}{\it
\begin{flushleft} \small
Talk presented at the 31$^\mathit{st}$ International Conference
on High Energy Physics, \\ 
24-31 July 2002, Amsterdam, The Netherlands.
\end{flushleft}
}\vspace*{21.6mm}
}

\author{P.A. Movilla Fern\'{a}ndez\address[MCSD]{
Max-Planck-Institut f\"ur Physik, F\"{o}hringer Ring 6, 
D-80805 M\"{u}nchen, Germany}}

\begin{document}

\begin{abstract}
  Re-analysed JADE data were used to determine \as\ at
  $\rs=14$-$44$\,GeV on the basis of resummed calculations for event
  shapes and hadronisation models tuned to LEP data.  The combined
  result is $\asmz=0.1194^{+0.0082}_{-0.0068}$ which is consistent
  with the world average.  Event shapes have also been used to test
  power corrections based on an analytical model and to verify the
  gauge structure of QCD.  The only non-perturbative parameter \az\ of
  the model was measured to $\aztwo=0.503\,^{+0.066}_{-0.045}$ and is
  found to be universal within the total errors.
\vspace{.75pc}
\end{abstract}

\maketitle

\section{INTRODUCTION}
The re-analysis of \epem\ annihilation data collected with the JADE
detector at the PETRA collider (1978-1986) has been shown to be a
valuable effort \cite{JADE-re1,JADE-re2,JADE-re2b,JADE-re3} since the
characteristic energy evolution of Quantum Chromodynamics (QCD)
becomes more manifest towards decreasing centre-of-mass energies
\rs. Recently, data at energies down to $\rs=14$\,GeV could be 
employed in state-of-the-art QCD studies due to the successful
resurrection of the original JADE software.

Since the PETRA shutdown, significant progress has been made in the
theoretical calculations of event shape observables.  In the
following, we present an \as\ analysis at $\rs=14$-$44$\,GeV based on
the most complete perturbative calculations for event
shapes~\cite{ERT,NLLA} available so far.  We included recently
analysed data in the energy region $\rs=14$-$22$\,GeV for which the
calculations are applied for the first time. Furthermore, power
corrections based on an analytical model by Dokshitzer, Marchesini,
and Webber~\cite{PC} (DMW) were investigated as a promising approach
to describe non-perturbative effects in event shapes.  Besides \as,
the model depends only on one additional free parameter. Also the
consistency of power corrections with the gauge structure of QCD was
tested.

\section{EVENT SHAPES}
From multihadronic data samples, the distributions of {\em thrust}
(\mbox{\thr}), {\em heavy jet mass} (\mh), {\em total} and {\em wide
jet broadening} (\bt\ and \bw), {\em \cp\ parameter} and the {\em
differential 2-jet rate} \yy\ in the Durham scheme are calculated
(cf.~\cite{JADE-re1}).  The data are corrected for the limited
acceptance and resolution of the detector and for initial state photon
radiation.  Since electroweak decays of the heavy b-hadrons fake hard
gluon radiation particularly at $\rs\le$ 22\,GeV, we take the
contribution $\epem\to\bbbar$ as an additional background to be
subtracted from the distributions.

The data have been used to assess the performance of various QCD event
generators tuned to LEP data at $\rs=\mz$ (cf.~\cite{tune}). The
parton shower and string fragmentation model implemented in \py/\jt\
is found to be well capable of describing event shapes down to
$14$\,GeV. The quality of the models \ar\ (colour dipole scheme) and
\hw\ (cluster fragmentation), however, is more moderate, and \cj\
(independent fragmentation) is clearly disfavoured.  Obviously, the
model parameters of these generators need a re-tune at lower \rs.

\section{DETERMINATION OF \as}
The determination of \as\ is based on a combination of an exact QCD
matrix element calculation \oaa~\cite{ERT} intended to describe the
3-jet region of phase space and a next-to-leading-logarithmic
approximation (NLLA) \cite{NLLA} valid in the 2-jet region where
multiple radiation of soft and collinear gluons from a system of two
hard back-to-back partons dominate.  We perform \chisq-fits of the
theoretical predictions corrected for hadronisation effects.  For the
main results, we use the \lnr-scheme~\cite{NLLA} for the perturbative
prediction with the renormalisation scale factor
\xmu~$\equiv\mu/\rs$~=~1 and \py\ for the estimation of
non-perturbative contributions.  We generally observe stable fits and
good agreement with the data at each \rs\ with $\chisqd\simeq$
0.2-2.0. In case of \bw, a significant excess of the theory over the
data in the 3-jet region of the distributions is present.

Experimental errors are under control for all data samples.
Expectedly, hadronisation uncertainties increase rapidly towards
14\,GeV. The individual results agree with each other within 1-2
standard deviations of the fit and experimental errors.  For each
\rs, the \as\ values from the six observables are combined using the
weighted mean method of Ref.\cite{JADE-re1} (Tab.~\ref{tab:as}).  The
total errors are dominated by higher order uncertainties. At 14 and
22\,GeV, hadronisation uncertainties are of the same order as the QCD
scale ambiguities.

\begin{table}[t]  \small
\caption{\label{tab:as}  \small
 Preliminary \as\ results from JADE.}
\renewcommand{\arraystretch}{1.2}
\renewcommand{\tabcolsep}{0.3pc}
\begin{tabular}{lccccc}  \hline 
\parbox{1.6pc}{\centering \rs \\ $[\gev]$}
   & \asrs\ & fit+exp. &  had. & hi. ord. & tot. \\ \hline
     14.0      &   $0.1704$ &   $\pm 0.0051$ & $ ^{+0.0141}_{-0.0136}$ 
        & $ ^{+0.0143}_{-0.0091}$ & $ ^{+0.0206} _{-0.0171}$ \\  
     22.0      &   $0.1513$ &   $\pm 0.0043$ & $\pm 0.0101 $ 
        & $ ^{+0.0101}_{-0.0065}$ & $ ^{+0.0144}_{-0.0121}$ \\  
     34.8      &   $0.1431$ &   $\pm 0.0019$ & $\pm 0.0073 $ 
        & $ ^{+0.0091}_{-0.0060}$ & $ ^{+0.0118}_{-0.0096}$ \\
     38.3      &   $0.1397$ &   $\pm 0.0040$      & $\pm 0.0054 $ 
        & $ ^{+0.0084}_{-0.0056}$ & $ ^{+0.0108}_{-0.0087}$ \\  
     43.8      &   $0.1306$ &   $\pm 0.0037$      & $\pm 0.0056 $ 
        & $ ^{+0.0068}_{-0.0044}$ & $ ^{+0.0096}_{-0.0080}$ \\  \hline
\end{tabular}\vspace*{-4.5mm}
\end{table}

\begin{figure}[t]  \vspace*{-0mm}
\includegraphics[width=.5\textwidth]{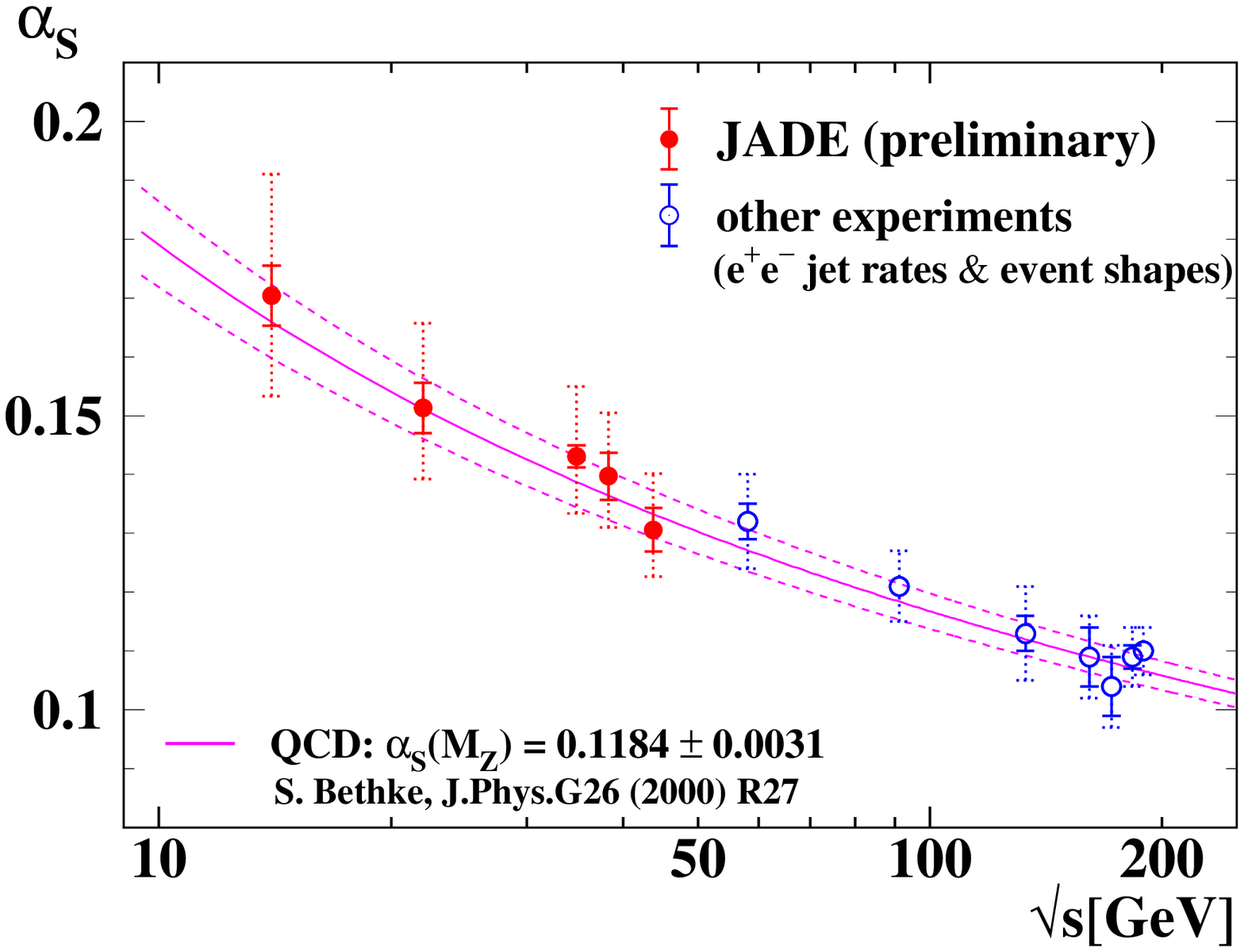} \vspace*{-13mm}
\caption{\label{fig:as} \small 
\asrs\ derived using \oaa+NLLA predictions for event shapes 
in \epem\ annihilation.} \vspace*{-4mm}
\end{figure}

The \as\ results obtained here and in similar analyses at higher
energies based on resummed event shapes (Fig.~\ref{fig:as}) agree well
with the QCD expectation for the running coupling~\cite{Bet00}. A
\chisq-fit taking statistical and experimental errors into account yields 
$\asmz=0.1213\pm0.0006$ with $\chisqd=8.3/11$.  Even considering the
total errors, the unphysical hypothesis $\as=$ const. is disfavoured
by a fit probability of $\approx10^{-5}$.

\section{TEST OF POWER CORRECTIONS}
The DMW model~\cite{PC} describes non-perturbative effects to event
shapes as contributions from gluon radiation at low energy scales,
assuming that the physical strong coupling \asmu\ remains finite in
the energy region around the Landau pole where simple perturbative
evolution of \as\ breaks down.  This leads to the introduction of a
parameter $\azmui=1/\mui\int_{0}^{\mui}\dd\mu\,\asmu$ that absorbs all
non-perturbative details of \asmu\ up to an arbitrary infrared
matching scale \mui.  The principle structure of power corrections is
a shift ${\calP}D_\var$ of the perturbative spectrum away from the
2-jet region, with $\calP\propto\az/\rs$ and $D_\var$ depending on
\var. In case of \bt\ and \bw, the shift is superimposed by a squeeze
$D_\var\propto\ln1/\var$. For mean values one obtains also an
additive correction.  The perturbative part is \oaa+NLLA for the
distributions and \oaa\ for the means.

The prediction, in particular the universality of \azmui, has been
tested by global fits to the hadron level data from this analysis and
from other experiments e.g. at LEP/SLC up to $\rs=$ 189\,GeV
(cf.~\cite{JADE-re2}), with \asmz\ and \aztwo\ as only free
parameters.  The major features of the distributions are reproduced
well, thus supporting the $1/\rs$ evolution of power corrections in
event shapes.  However, we observe discrepancies for the distributions
of the less inclusive quantities \mh\ and \bw\ in particular at
$\rs<\mz$.

For \yy\ the leading power correction is known to be quadratic in
$1/\rs$.  This expectation has been verified by means of the new JADE
distributions at $\rs=14$ and 22\,GeV using a simple additive power
correction ansatz.

\begin{figure}[t]\vspace*{-0mm}
\centerline{
\includegraphics[width=.3\textwidth]{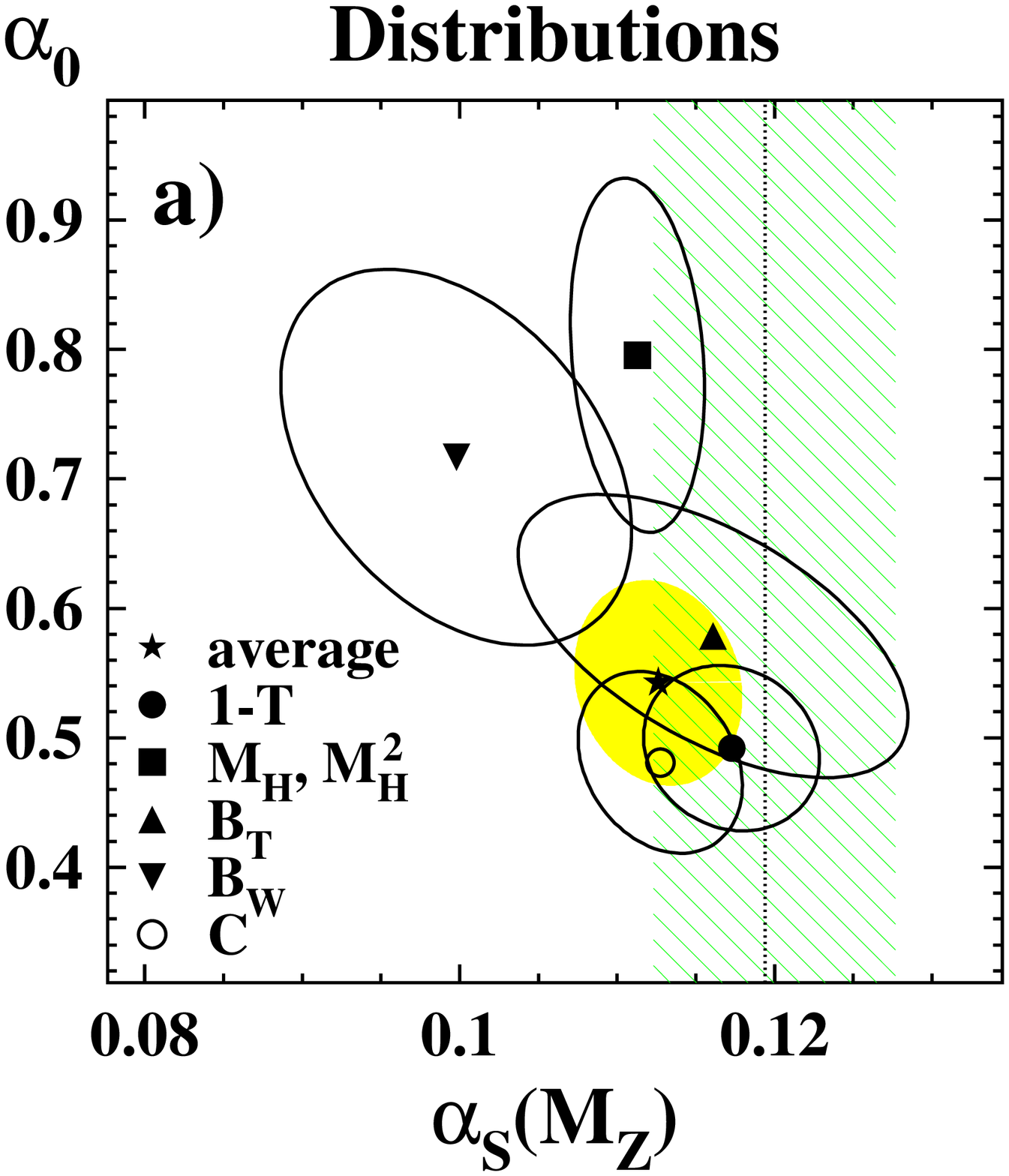}
\hspace*{-10mm}
\includegraphics[width=.3\textwidth]{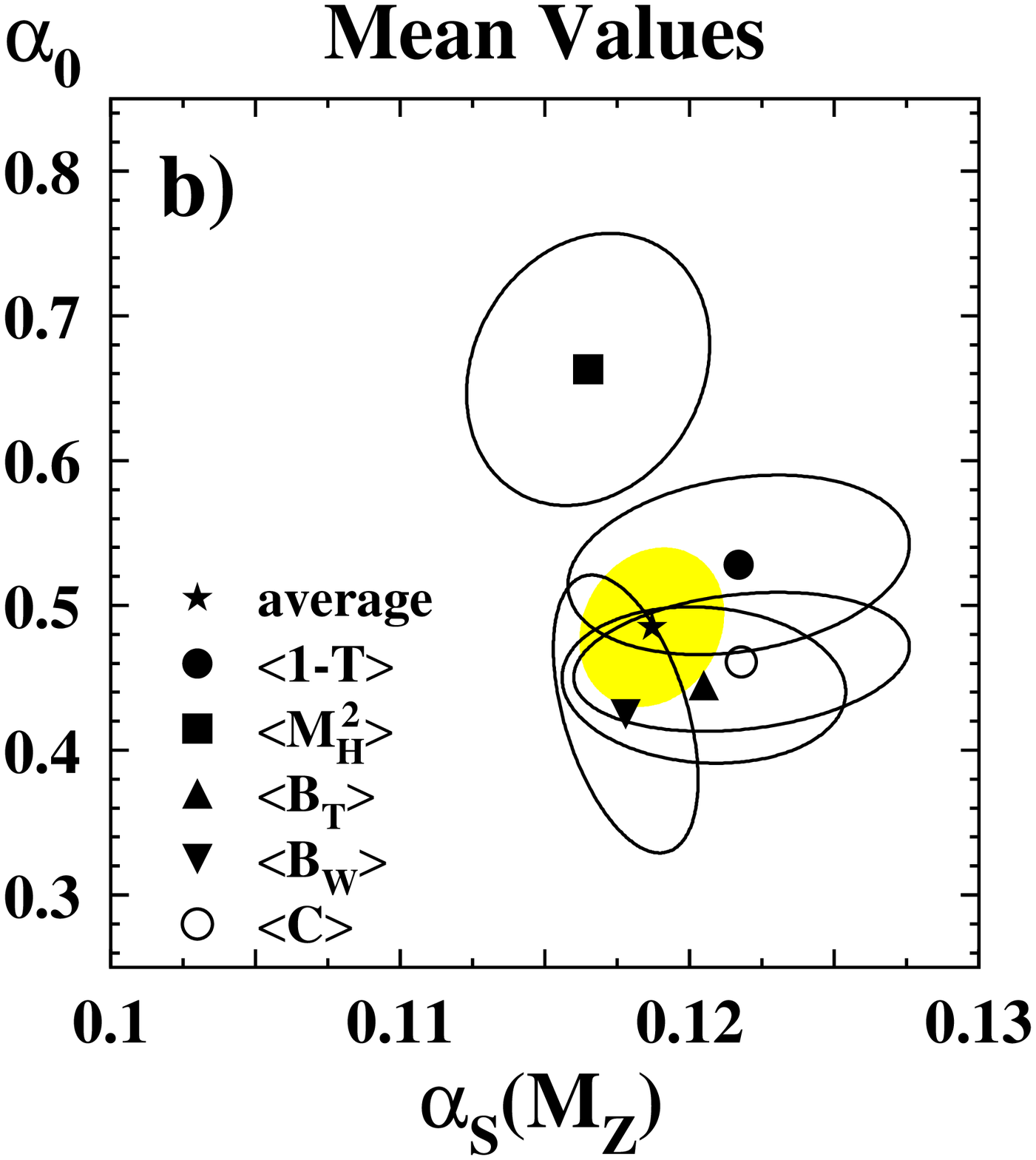}
}\vspace*{-8mm}
\caption{\label{fig:PC}\small 
\asmz\ and \aztwo\ and one standard deviation errors of DMW-fits 
to distributions a) and means b).  The hatched band represents the
combined \asmz\ derived from the ``conventional'' analysis.
}\vspace*{-5mm}
\end{figure}

As shown in Fig.~\ref{fig:PC}, there is a reasonable agreement between
the individual results within the total uncertainties. However, the
\az\ results from the \mh\ and \bw\  distributions are large compared 
to the results from the other observables.  This observation may be
related to the non-inclusiveness of these variables. The \as\ values
from power corrections to the distributions are systematically smaller
than the results based on MC corrections.  This is due to the
different amounts of squeeze of the perturbative spectrum predicted by
both types of model, particularly in case of \bw.

Combining the results for the mean values and the distributions taking
correlations between the systematic errors into account yields
$\asmz=0.1175^{+0.0031}_{-0.0021}$ and
$\aztwo=0.503^{+0.066}_{-0.045}$. The scatter of the \az\ values is
mostly covered by the theoretical uncertainty of the Milan
factor~\cite{PC}.

\section{STUDY OF QCD COLOUR FACTORS}
The DMW ansatz has been exploited to extract the QCD colour factors
$\CA$, $\CF$, and $\nf$~\cite{JADE-re2b}. Various global fits to the
event shape spectra trying alternative sets of the free model
parameters support the $SU(3)$ symmetry group. The most stable and
precise measurements are provided by \thr\ and \cp.  Combining the
corresponding results for these variables with \aztwo\ and \nf\ fixed,
and \asmz, $\CF$, and $\CA$ free, one finds $\CF=1.29\pm0.18$ and
$\CA=2.84\pm0.24$. This is in good agreement with the QCD expectation
while some other gauge symmetry groups are excluded.

\section{CONCLUSIONS}
Resummed QCD theory combined with LEP tuned hadronisation models fits
event shape data well down to $\rs=14$\,GeV and allow consistent
determinations of \as. The combined result evolved to the $\znull$
mass scale is $\asmz=0.1194^{+0.0082}_{-0.0068}$ which is
substantially more precise than former PETRA measurements and also in
good agreement with the world average value \cite{Bet00}.

Power corrections $\propto1/\rs$ generally reproduce the overall event
shape spectra, except for the distributions of the less inclusive
variables (\mh\ and \bw) at $\rs<\mz$. The results for \az\ support
the DMW prediction of universality within 25\%. Using power
corrections, the gauge structure of QCD has been verified with
uncertainties competitive e.g. with traditional 4-jet angular
correlation analyses. Potential biases from hadronisation models are
reduced within this approach.

Thus, exploiting JADE data significantly improves the verification of
QCD on the basis of \epem\ annihilation.

\end{document}